\documentstyle[aps,prl,twocolumn,epsfig,amsmath,amssymb]{revtex}

\begin{document}
\twocolumn[\hsize\textwidth
\columnwidth\hsize\csname
@twocolumnfalse\endcsname

\title{Coarsening Dynamics of a Quasi One-dimensional Driven Lattice Gas}
\author{J.T. Mettetal, B. Schmittmann, and R.K.P. Zia}
\address{Center for Stochastic Processes in Science and Engineering, Physics
Department \\
Virginia Polytechnic Institute and State University, Blacksburg, VA
24061-0435, USA}
\date{\today }
\maketitle

\begin{abstract}
We study domain growth properties of two species of particles executing
biased diffusion on a half-filled square lattice, consisting of just two
lanes. Driven in opposite directions by an external ``electric'' field, the
particles form clusters due to steric hindrance. While strictly
one-dimensional systems remain disordered, clusters in our ``quasi 1D'' case
grow until only a single macroscopic cluster survives. In the coarsening
regime, the average cluster size increases $\sim t^{0.6}$, significantly
faster than in purely diffusion-controlled systems. Remarkably, however, the
cluster size distribution displays dynamic scaling, following a form
consistent with a diffusion-limited growth mechanism.
\end{abstract}
\vspace{-0.5cm}
]

\narrowtext

{\em Introduction. }The dynamics of a system undergoing phase segregation
when quenched below the transition temperature has been of interest for many
years. Starting from a homogeneous state, domains of the co-existing phases
form and grow. Nearly all existing studies of such coarsening processes are
devoted to systems subjected to dynamical rules which eventually take them
to {\em equilibrium} states \cite{GB}. Our interest is coarsening in systems
which are evolving toward {\em nonequilibrium }steady states (NESS): The
underlying dynamics of our systems violates detailed balance and time
reversal invariance. In particular, our main focus will be the remarkable
characteristic of dynamic scaling. In this letter, we report the presence of
this behavior in a simple model of biased diffusion of two species. In stark
contrast to similar phenomena in other systems where the growth exponent is
typically 1/2 or 1/3 \cite{MG,CB,ln}, we find it to be at least $0.6$! This
large, anomalous value rules out diffusion-controlled growth mechanisms.
Surprisingly, however, the scaling function for the cluster-size
distribution resembles that of diffusion-dominated growth \cite{CB,BBD}
quite well. Since our system is essentially one-dimensional, we may also
exploit a reduced description, in terms of coalescing random walkers on a
line. To account for the faster growth, we introduce {\em interactions}
between neighboring walkers. Apart from small deviations, the
``cluster-size'' distributions in this picture are in good agreement with
the data. In the following, we briefly describe the model, present the data
from Monte Carlo (MC) simulations, discuss the coalescence picture, and
conclude with some open questions. More details will be left to \cite
{MSZ-long}.

{\em Biased Diffusion of Two Species.} Motivated by the physics of fast
ionic conductors \cite{FIC}, Katz, {\em et.al.} \cite{KLS} generalized the
well-known Ising lattice gas to include a uniform bias affecting all
particles. With periodic boundary conditions, the system settles into a NESS
rather than a simple equilibrium state. In addition to a non-trivial
particle current, many novel collective phenomena emerge, some of which
remain unexplained \cite{rev}. A natural extension of such ``driven
diffusive systems'' involves {\em two }species. When these are driven in
opposite directions, the system displays a phase transition in $d=2$, even
if the {\em only }interparticle interaction is an excluded volume constraint 
\cite{shz,ksz1}. In $d=1$, however, one can prove \cite{SG} that phase
transitions cannot occur. Since such a system consists of particles on a
line, it may be used to model traffic flow \cite{traffic}, and we will use
the term ``single lane'' for this $d=1$ case. Remarkably, when a second
``lane'' is introduced, the system again displays a phase transition \cite
{ksz2}! We briefly describe this study.

A fully periodic $2\times L$ square lattice is randomly filled with equal
numbers of two types of particles (labeled by ``charge'' $+$ or $-$),
subject to an excluded volume constraint. When not driven, the particles
diffuse according to the following rules. Two nearest neighbor (NN) sites
are chosen at random, and attempts to exchange their contents are (a) always
allowed for particle-hole pairs and (b) accepted with probability $\gamma $
for particle-particle pairs. Under this dynamics, the two species are just
diffusing randomly. The system obviously remains homogeneous, with trivial
collective properties. Next, we impose an external ``field'' which drives
the two species aroung the ``ring'' in {\em opposite} directions, by simply
forbidding all exchanges which result in, say, $+/-$ particles moving
clockwise/anticlockwise. In \cite{ksz2}, $\gamma $ is fixed at $0.1$ while
systems up to $L=10^{4}$ are evolved for as many as $4\times 10^{6}$ MC
steps ($1$ MCS corresponds to $2L$ exchange attempts). For early times
(10-20 MCS), small blockages form everywhere, due to the mutual obstruction
of the opposing species. These clusters coarsen until, at late stages when
NESS is essentially established, only a single macroscopic cluster remains.
The characteristics of this cluster are notable. (a) It contains almost no
holes. (b) It exhibits a non-trivial ``charge'' profile. (c) Its size
(number of particles connected as NN's) is gaussian distributed around $%
0.94L $, so that its ``length'' is about $L/2$. (d) A low-density region
spans the remainder of the system, consisting of particles (``travellers'')
which leak out of the cluster at one end and later rejoin it at the other.
By contrast, for the ``single lane'' case, the system remains homogeneous,
and the cluster size distribution is known exactly \cite{SG} to decay
exponentially (with weak $1/L$ corrections). In this sense, we believe that
it is justifiable to use the terms ``long range order'' and ``phase
segregation'' to characterize the two-lane system in steady state. Of
course, if the overall particle density is reduced sufficiently, there must
be a phase transition to re-establish homogeneity. In an effort to
understand why a two-lane system orders while the single-lane version
remains disordered, we turn to a study of {\em how} long-range order builds
up in the latter case.

{\em Simulations, Characterizations, and Results.} To study the time
dependence of our stochastic system, we carry out many runs with the same
random initial distribution of particles. Typically, data from $100-1000\,$%
runs are used to build histograms. In addition, with finite size effects in
mind, we exploit a range of system sizes, from $2\times 100$ to $2\times
1000 $. At half filling, local blockages appear everywhere soon after the
run starts, a phenomenon reminiscent of homogeneous nucleation. Though
small, these blockages share some characteristics of the terminal cluster,
having few holes inside and growing through accretion of particles from the
inter-cluster region. In a relatively short time ($\sim $100 MCS), a rough
balance between accretion and loss is established so that both the density
of travellers and their total number remain relatively constant for the rest
of the run. The associated values, $\sim 5\%$ and $\sim 0.06L$ respectively,
can be understood qualitatively \cite{ksz2}. Thus, we may characterize this
initial stage as one in which {\em local} densities quickly reach their
(approximate) final values. For later, intermediate times (the ``growth''
regime), the larger clusters grow at the expense of smaller ones, a
mechanism reminiscent of Lifshitz-Slyosov growth. However, the similarity is
deceptive, as we will show next.

To characterize domain growth in our model, we monitor the cluster size
distribution since it carries very detailed information. The most natural
definition of the size of a cluster, $s$, is just the number of particles
(regardless of their charge) connected via NN bonds. We collect data at
various times, $t,$ to construct histograms of such cluster sizes: $\tilde{p}%
(s,t)$. Since this distribution is sensitive to both the growing clusters
and the travellers, we refer to it as the {\em microscopic} cluster
distribution. Clearly, $\tilde{p}$ is not conserved, decreasing when
clusters merge, etc. Instead, we consider $\,p\left( s,t\right) \equiv s%
\tilde{p}\left( s,t\right) $, which, counting just the total number of
particles, {\em is }conserved. Known as the ``residence distribution'' in
percolation studies, $p\left( s,t\right) $ is proportional to the
probability of any site belonging to a cluster of size $s$ at time $t$. For
the $2\times L$ case, $p$ displays two peaks from an early time, one at $s=1$
(followed by a sharp drop to nearly zero) and another moving to higher $s$
with $t$. As mentioned above, the first is relatively constant, so that we
focus only on the behavior of the second peak. To this end, we devise a {\em %
coarse-grained }(CG) description of the clusters, which has the added
advantage of tolerating the occasional hole in a cluster. For any
configuration on the $2\times L$ lattice, we construct another one, with
occupation numbers $0$ or $1$on a {\em line} of $L$ sites, as follows. At
each site $i$, we assign $0$ if there are five or less particles in the ten
sites around the $i^{th}$ column of the original lattice; and $1$ otherwise
(cf.~Fig.~1.)

\begin{center}
\begin{eqnarray*}
&& 
\begin{tabular}{|p{0.15in}|p{0.15in}|p{0.15in}|p{0.15in}|p{0.15in}|p{0.15in}|p{0.15in}|p{0.15in}|p{0.15in}|p{0.15in}|p{0.15in}|p{0.15in}|}
\hline
& $-$ &  & $+$ & $+$ & $+$ & $-$ & $-$ & $-$ &  &  &  \\ \hline
$+$ &  & $+$ & $+$ & $-$ & $+$ & $-$ & $+$ & $-$ &  & $-$ &  \\ \hline
\end{tabular}
\\
&& 
\begin{tabular}{|p{0.15in}|p{0.15in}|p{0.15in}|p{0.15in}|p{0.15in}|p{0.15in}|p{0.15in}|p{0.15in}|p{0.15in}|p{0.15in}|p{0.15in}|p{0.15in}|}
\hline
0 & 0 & 1 & 1 & 1 & 1 & 1 & 1 & 1 & 0 & 0 & 0 \\ \hline
\end{tabular}
\end{eqnarray*}
\end{center}

\vspace{-1.cm} 
\begin{figure}[tbp]
\caption{Illustration of the coarse-graining procedure for a configuration
in a $L=12$ {\em periodic }lattice. An example of the microscopic
configuration is shown with $\pm $ particles and holes. The CG version shows
only $\left\{ 0,1\right\} $'s.}
\end{figure}
\vspace{-.2cm}

\noindent In the CG configuration, a cluster is simply a consecutive
sequence of $1$'s and its ``size'' ($\ell $) is just the length of this
string. Histograms for these cluster sizes, $\tilde{p}(\ell ,t)$, can be
compiled easily. Note that clusters in the traveller region are usually
quite small ($s\leq 5$), so that $\tilde{p}(\ell ,t)$ has only one peak,
associated with the growing clusters. Again, we consider the residence
distribution: $p\left( \ell ,t\right) \equiv \ell \tilde{p}\left( \ell
,t\right) $, from which a natural definition of an average cluster length
arises: $\bar{\ell}_{C}\left( t\right) \equiv \sum_{\ell }\ell \cdot p(\ell
,t)/\sum_{\ell }p(\ell ,t)$. In Fig.~2, we present a log-log plot of this
quantity for various $L$ (solid points). Clearly, there is good data
collapse in the growth regimes. For the two smaller systems, our runs are
long enough for saturation ($\lim_{t\rightarrow \infty }\bar{\ell}%
_{C}(t)\simeq 0.47L$) to occur. More significantly, the effective power in
the growth (i.e., $d\ln \bar{\ell}_{C}/d\ln t$) appears to be still
increasing for the largest system at the latest times. In the figure, we
provide two straight lines, corresponding to $t^{1/2}$ and $t^{2/3}$, to
guide the eye. From these lines, we estimate that the effective exponent is
about $0.6$ at the latest stage. Unless a mysterious slow-down sets in at
even later times, we are confident that the power of $1/2$ is ruled out.
Finally, we turn to the distributions and, to check for dynamic scaling,
construct the normalized scaled form, $f\left( x,t\right) \equiv \bar{\ell}%
_{C}\left( t\right) p(\ell ,t)$ with $x\equiv \ell /\bar{\ell}_{C}\left(
t\right) $. As we see in Fig.~3a, $f\left( x,t\right) $ is largely
independent of $t$ (and $L$) in the growth regime. Thus, we conclude that,
despite the possible absence of pure power law growth, the distribution
displays, remarkably, dynamic scaling. Further, the scaled distribution is
well approximated by the solid line in Fig.~3a, a distribution (cf.~Eqn.~(%
\ref{BBD}) below) well known from diffusion-limited coalescence \cite{CB,BBD}%
. This suggests casting our MC dynamics in terms of an effective model,
involving coalescing random walkers (RW).

{\em Coalescing random walkers in $d=1$}. Since the density of travellers
is relatively constant, a good macroscopic picture focuses on the sizes
(lengths) of the clusters only, with the main dynamics being particle
exchange between neighboring clusters. To describe this sequence of
clusters, points (``walkers'') are placed on a ring of $N$ sites, with
spacings corresponding to the cluster-lengths \cite{CB}. Particle exchanges
translate into walkers taking steps. When two walkers meet, they 
``coalesce,'' representing the disappearance of a cluster. The {\em number}
of walkers, $n$, maps into the number of clusters, while $N$ represents the
sum of all cluster lengths. At the end, there is only one walker,
i.e., one macroscopic cluster. For simplicity, we set $N=L/2$ which well
approximates the observed $0.47L$. In our simulations, RW time ($\tilde{t}$)
increases by $1$ when every walker has had, on average, a chance to make one 
step.

\begin{figure}[tb]
\input{epsf} 
\vspace{-.5cm} 
\hspace{2.0cm} 
\begin{minipage}{.4\textwidth}
    \epsfxsize = 1.1\textwidth \epsfysize = .9\textwidth 
  \epsfbox{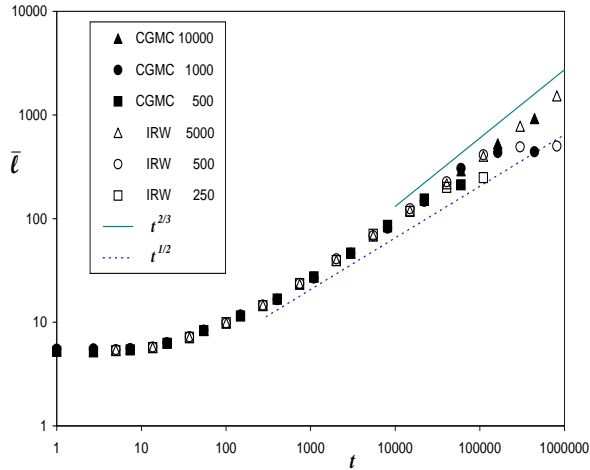} 
  \end{minipage}
\caption{Coarse-grained MC (solid symbols) and interacting random walk (open
symbols) data for the average cluster size vs time, for $L=10000,1000$ and $%
500$.}
\end{figure}
\vspace{-.3cm}

If the random walkers were free (apart from coalescence), the problem is
exactly soluble in $d=1$ \cite{BBD,CB}. Denoted by $\tilde{P}_0(\ell ,\tilde{%
t})$, the (unnormalized) frequency of finding two adjacent walkers separated
by a distance $\ell $, at time $\tilde{t}$, satisfies a diffusion equation
in the continuum limit. With absorbing boundary condition at $\ell =0$ to
model coalescence, the solution is standard. To compare with the MC data,
consider the {\em normalized} ``residence'' distribution, $P_0(\ell ,\tilde{t%
})\equiv \ell \tilde{P}_0(\ell ,\tilde{t})$. Its first moment, $\bar{\ell}_0$%
, should correspond to $\bar{\ell}_C$. Clearly, $\bar{\ell}_0$ grows
diffusively: $\bar{\ell}_0(\tilde{t})\propto \tilde{t}^{1/2}$. Further,
dynamic scaling prevails (in the limit $\tilde{t}\rightarrow \infty $, $\ell
\rightarrow \infty $ at fixed $x\equiv \ell /\bar{\ell}_0(\tilde{t})$) : 
\begin{equation}
\bar{\ell}_0(\tilde{t})P_0(\ell ,\tilde{t})\equiv F_0(x)=\frac{32}{\pi ^2}%
x^2\exp (-\frac 4\pi x^2)\text{ },  \label{BBD}
\end{equation}
which serves as a good benchmark (solid line in Figs. 3a and b) for scaling
plots of the data. Unfortunately, the growth exponent ($1/2$) of this simple model 
is unequivocally ruled out by our data (Fig. 2). Its deficiencies can
be traced, by monitoring the evolution of our clusters, to the rapid, {\em %
systematic} disintegration of small ones. To model this bias towards larger
clusters, we introduce a model of interacting random walkers (IRWs), where
neighboring walkers experience stronger {\em attraction} with decreasing
separation. Specifically, consider a walker and its two nearest neighbors.
Letting $\ell _R$ ($\ell _L$) be the distance to its right (left) neighbor,
and $q_R,$ $q_L,$ $q_S(=1-q_R-q_L)$ be the probabilities for moving to the
right, left, and staying, respectively, we choose: 
\begin{equation}
\frac{q_R}{q_S}=1+\left( \frac C{\ell _R}\right) ^2\text{ \quad and\quad }%
\frac{q_L}{q_S}=1+\left( \frac C{\ell _L}\right) ^2  \label{IRW_rates}
\end{equation}
where $C$ is an amplitude giving the best fit to the data.

\begin{figure}[t]
\input{epsf}  
\vspace{-.5cm}
\hspace{2.0cm} 
\begin{minipage}{.4\textwidth}
    \epsfxsize = 1.1\textwidth \epsfysize = 1.1\textwidth 
  \epsfbox{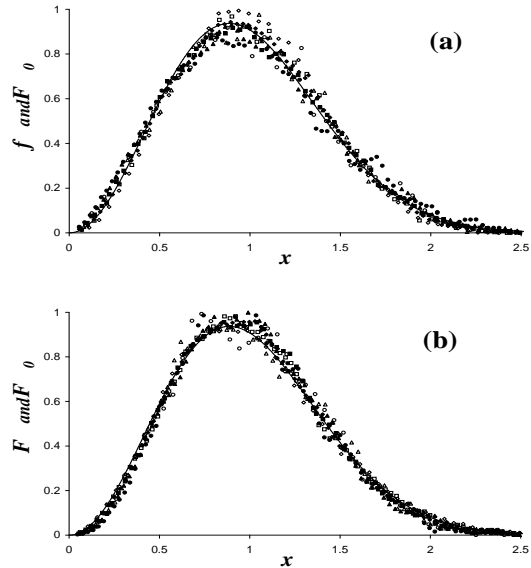} 
  \end{minipage}
\caption{Scaled residence distributions for CG MC, $f(x)$, and IRW, $F(x)$. For
comparison, the ``benchmark,'' $F_0$, is shown
as a solid line in both plots. In (a), $\blacklozenge ,~\blacksquare ,
~\blacktriangle$ and $\bullet$ are data from a $2\times 2000$ lattice at $%
t=1096,2981,8103$, and $2207$, respectively. Open symbols are
data from $2\times 500$ at $t=403,1096,2981,$ and $8103$. In (b), this set
of solid (open) symbols correspond to $N=1000\left( 500\right) $, at scaled
times.}
\end{figure}
\vspace{-.2cm}

Some comments are in order. {\em (i)} To motivate the $\ell ^{-2}$%
-dependence, we invoke the mean-field theory for the two-species model \cite
{ksz1}. Considering the continuity equations for mass and charge density for
the traveller region and the interior of a cluster separately, and matching
the two regions via a shock condition \cite{Bonn}, we find the steady-state
charge current through a cluster of length $\ell >>1$ to be $j(\ell
)=(\gamma /2)[1+O(\ell ^{-2})]$ \cite{MSZ-long}. Symmetry dictates that the
mass currents of the $+/-$ species are equal and opposite. Thus, for neighboring
clusters with $\ell _R$ and $\ell _L$, say, the imbalance in the currents
would be $\sim \left( \ell _R^{-2}-\ell _L^{-2}\right) $. In turn, this
current difference leads to one cluster gaining at the expense of its
neighbor.{\em (ii)} To account only for a bias, $q_S\equiv 0$ would suffice. Our
choice, allowing a walker between small clusters to be more mobile, models
the faster dynamics observed in that case. {\em (iii)} If $C$ were a simple
constant, the behavior at late times (with fewer but larger clusters) would
again be dominated by $\ell $-independent rates, leading to $\tilde{t}^{1/2}$
growth. As a remedy, we must supply a scale, namely $C$, against which to 
measure ``small'' clusters. 
Thus, $C$ is chosen to depend on the average size of clusters,
i.e., $N/n$. Finding $C\simeq \left( 0.06N/n\right) ^2$ to fit best, we 
provide the following heuristic argument in its favor. On the average, an $%
\ell $-cluster will find itself sandwiched between two sections of
travellers with length $N/n$. When $\ell \ll N/n$, the overall density in
that region will be too low to support inhomogeneities and the cluster
should decay quickly. As for the value $0.06$, we note that this is also the
traveller density, but can offer no insight into this coincidence.

This choice for $C$ renders our rates time-dependent (through $n(%
\tilde{t})$), so that the exact solution of this dynamics remains elusive.
Fortunately, our IRW is easily simulated. At time $\tilde{t}=0$,
walkers are distributed randomly over the ring, with an initial density of $%
0.4$, and then updated according to (\ref{IRW_rates}). We measure the
normalized {\em residence distribution} $P(\ell ,\tilde{t})$ and its first
moment, $\bar{\ell}_I(\tilde{t})$. Adjusting a single scale parameter to match $%
\tilde{t}$ to $t$, we plot (in Fig. 2) both $\bar{\ell}_C$ and $\bar{\ell}_I$%
, for a range of system sizes. The success of our effective model is quite
staggering: the two measures for $\bar{\ell}$ trace out the same curve for 
{\em over four decades}. The small differences observed in the steady state
values are easily understood: while $\bar{\ell}_I(t)$ approaches a final
value of $L/2$, the corresponding limit of $\bar{\ell}_C(t)$ is only $\sim
0.47L.$ Finding $P(\ell ,\tilde{t})$ to be qualitatively similar to its
counterpart from MC, we construct the scaling plot: $F(x)\equiv \bar{\ell}_I(%
\tilde{t})P(\ell ,\tilde{t})$ vs $x\equiv \ell /\bar{\ell}_I(\tilde{t})$
(Fig. 3b). Within statistical noise, the data collapse is quite
satisfactory. Most remarkably, this $F(x)$ also closely follows $F_0\left(
x\right) $ of the free RW!

There are, however, small deviations from scaling, revealed only by {\em %
close} inspection. We first note that, in both cases, the late-time data
tend to deviate systematically from $F_0(x)$. In Fig. 3a, the data lie
slightly above $F_0$ for $x\leq 0.5$, and slightly below $F_0$ for $x$ near
its maximum. In Fig. 3b, the trend is reversed. For $x\geq 1$, there are no
perceptible deviations in either case. While we cannot offer any rigorous
understanding, it is conceivable that the lengthening domains of travellers
- which are neglected in the IRW dynamics - play a role here. Another
possible mechanism is described in \cite{SKR}. Much more remarkable,
however, is how close both $f$ and $F$ follow Eqn. (\ref{BBD}), i.e., 
the free RW scaling function,
even though neither $\bar{\ell}_C$ and $\bar{\ell}_I$ are consistent
with the diffusion equation leading to $F_0$! It appears as if the
complexity of our full dynamics can be absorbed into a single
``renormalization'' of time, $t\propto \tau ^\sigma $, such that simple
diffusion-controlled coalescence re-emerges in $\tau $. Matching $%
t^{1/z}\sim \bar{\ell}\sim \tau ^{1/2}$ then requires $\sigma =z/2$.
Clearly, formalizing this scenario poses a serious theoretical challenge.

{\em Summary and Outlook. }To conclude, we summarize our findings and end
with some speculations. We consider a lattice gas with two species of
particles driven in opposite directions on a ``single-lane road'' vs. a
``two-lane highway.'' Allowing a small amount of particle-particle exchange,
both systems are ergodic but display drastically different final states:\
While the one-lane system remains homogeneous, the two-lane case exhibits a
macroscopic cluster which scales with system size. In an effort to probe the
surprisingly different outcomes, we study the growth of clusters in the
two-lane case in some detail, measuring the average cluster size, $\bar{\ell}%
(t)$, and the full residence distribution. We discover that $\bar{\ell}(t)$
grows {\em much faster} than typical domains in similar models. At late
times, the growth law may be characterized by an effective exponent of about 
$0.6$. We end with a speculative note. In ordinary Lifshitz Slyosov growth,
which leads to the power $1/3$, the mechanism for transport (of material
from smaller clusters to larger ones) is diffusion. Here, particles suffer 
{\em biased }diffusion, so that the motion is ballistic instead. Can the
difference of a factor of two between these different mechanisms transform
the power $1/3$ to a $2/3$? Needless to say, to arrive at definitive
conclusions, we need not only a deeper analytic understanding but also a
better Monte Carlo study of this simple, yet intriguing, model system.

{\em Acknowledgments.} We thank A. Vasudevan and G. Korniss for helpful
discussions. The research is supported in part by the 
National Science Foundation through grant DMR-0088451.
One of us (JTM) acknowledges NSF support via an REU supplement.

\end{document}